**Falling non-harmonic Slinkys**

Paul Hatchell, Independent researcher, Katy, TX

**Abstract**

Slinkys that start from a stretched equilibrium position supported at the top and then released to fall under the influence of gravity exhibit the interesting behavior that the bottom of the slinky does not move until the collapsing top of the Slinky reaches the bottom. In this paper, we examine this problem using numerical methods to investigate whether this property holds for generalizations of the slinky physics such as changing the restoring force from the traditional Hooke's law or considering random and non-uniform distributions of masses. For restoring forces, F, of the type $F = kx^p$, where x represents the spring displacement and k the generalized spring constant, it is found that when $p > 0$, the bottom-doesn't-move property holds, but when $p < 0$, the model shows complicated collapse patterns that in some cases depend on whether the number of modeled masses is even or odd.

**Introduction**

The falling slinky demonstration is easily performed in a classroom with an outcome that comes as a surprise and delight to many. The problem was highlighted as a "physics trick of the month" by Martin Gardner[1] (2000) and has been featured in several popular science videos such as those published by Veritasium[2] (2012) that have received millions of views. The main point of interest in these demonstrations is that when a slinky is released to fall from a stretched equilibrium position it is observed that the bottom of the slinky does not move until the collapsing top of the slinky reaches the bottom. This so-called "slinky levitation" also occurs when the slinky is modified such as by attaching a point mass to the bottom of the slinky (See for example the Veritasium[2] videos and Dubinov and Dubinova[3] (2021) )

Solutions to the falling slinky problem have been presented by many authors including Calkin[4] (1993) , Unruh[5] (2011) and Cross and Wheatland[6] (2012) who model the slinky behavior using an analytic approach based on continuum mechanics and partial differential equations. Another set of modelling solutions have been presented by Graham[7](2001), Plaut et al [8](2015), and Vanderbei [9](2017) that model the slinky as a finite chain of discrete point masses connected by massless springs and using basic laws of mechanics. Vanderbei[9] argues that in the limit of infinite point masses that the discrete case should approximate a continuous slinky although this is not precisely true because of considerations presented by Cross and Wheatland[6] concerning properties of a real tension springs having collapsed turns of finite thickness.

In the analytical approach, solutions are readily obtained because of the special properties of Hooke's law with regards to the equations of motion. Specifically, when the restoring force is proportional to displacement, y, so that F=-ky, the differential equations in the center of mass system (where gravity can be ignored) have the form mÿ+ky=0, where m represents mass. These types of differential equations are easily solved using harmonic functions which have the property of proportionality between the harmonic function (such as sin and cos) and its second derivative and the problem is reduced to solving a 2nd order algebraic equation.

For non-Hooke's law forces or when non-uniform mass distributions are used, the analytical approach is difficult to generalize and numerical methods using discrete point masses are better adapted to exploring a wide range of variations of the slinky physics and should be accessible to students at many levels who can explore this problem using straightforward computational methods. It is the approach that is taken here.

In this paper we will investigate whether the bottom-doesn't-move property holds for variations of the slinky physics such changing the mass distribution and deviations from the Hooke's law model that is used in previous modeling studies. In the next session, we set up the discrete mass problem and initial conditions for the modified slinky. In the following section solutions to this are computed numerically as the number of point masses is varied, the mass distribution is varied, and what happens with non-Hooke's law forces.

**Initial conditions**

We approximate the slinky as starting with a vertical chain of N point masses coupled by massless generalized springs. The goal is to model what happens with variations in point mass distributions as well as deviations from Hooke's law restoring forces in the springs. For two masses separated by a positive distance, y, we consider forces, F, of the type

$$F(y) = -ky^p, \qquad (1)$$

where the exponent, p, is allowed to vary. The case p=1 corresponds to Hooke's law, but there are many exponents of interest such as p=-2 (electrostatic and gravity) or p=-3 (magnetic dipoles). The minus sign in eq. (1) indicates that this is a restoring force. Note that the units of the generalized spring constant, k, are $Force/Length^p$.

Figure 1 shows the geometry used in the model. There are N point masses connected by N-1 springs. The mass at each point is $m_i$ and its vertical position as a function of time, t, is $y_i(t)$. The values of $y_i$ are ordered such that $y_N$ has the largest value and represents the top of the slinky, and $y_1$ corresponds to the smallest value and represents the bottom of the slinky. Gravity acts downwards in this model.

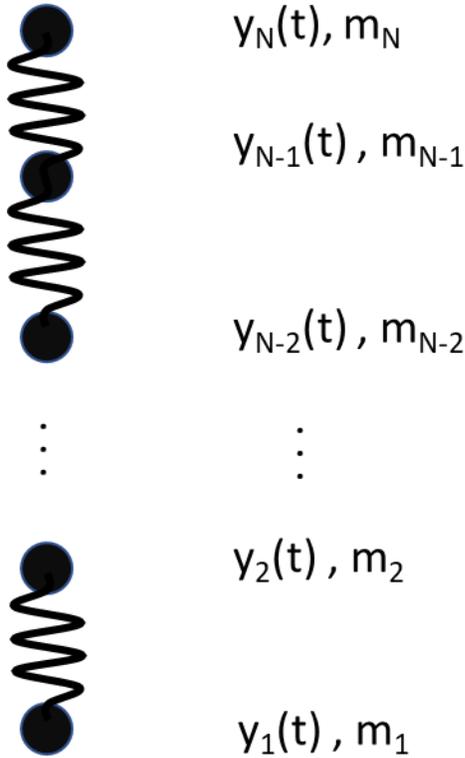

Figure 1: Geometry of the discrete mass slinky model with N masses have mass $m_i$. The positions as a function of time are denoted by $y_i(t)$.

The slinky starts prior to its release (at time t=0) from an equilibrium position where it is supported from the top and allowed to hang freely at rest. In equilibrium, the restoring force in each spring needs to counteract the force due to gravity of all the masses that are positioned below the spring. Balancing these forces gives the result for the relative positions between point masses:

$$y_{i+1}(0) - y_i(0) = \left[\frac{g \sum_1^i m_i}{k}\right]^{1/p}, \tag{2}$$

where g is the acceleration due to gravity. There is an important scaling relationship in this equation that we take advantage of. The masses, $m_i$, and k appear as a ratio. If we scale each mass and the spring constant by the same quantity the starting positions are left unchanged.

The crucial parameter that controls the time evolution of the slinky is its starting equilibrium length, $L_S$. In order to easily compare models with different values of $m_i$, p, and N we will define a impose two requirements and define a "standard $L_S$ slinky":

1) The equilibrium length of the slinky is $L_S = y_N(0) - y_1(0)$, (3a)
2) The equilibrium position of the slinky center of mass is at y=0. (3b)

The first requirement of a fixed starting length establishes the value of the spring constant k and the second requirement fixes the absolute position of the all the point masses. One way to implement this numerically is to set k=1, $y_1=0$ and then use eq. (2) to determine the starting values of $y_2$ through $y_N$. A corrected k value is then determined by comparing the computed slinky length $y_N(0)-y_1(0)$ with the required length, $L_S$. The correct k value is then given by $k=[L_S/(y_N(0)-y_1(0))]^p$. Once the value of k is known, the correct starting positions can be determined and translated to place the center of mass at y=0.

For the standard $L_S$ slinky, it is not necessary to specify the total mass of all the point elements because of the scaling relationship described previously. All that is necessary is to describe their distribution. The simplest case occurs when all elements have identical masses, $m_i$=constant, and we will consider other distributions such as random masses, top loaded masses and bottom loaded masses later.

Figure 2 shows the starting equilibrium positions for standard $L_S$ slinkys having uniform masses and for values of N=3, 5, 10, 20 and where the p value ranges from -4 <= p <= 4. Note that the result is undefined when p=0 (except for the case N=2). $L_S$ is chosen to be 2.0 m for this and all future examples and note that if we change Ls the starting positions scale proportionally. For N=2 (not shown), the starting positions are +1.0 and -1.0 m. For values of N > 2, there is a discontinuity at p=0. For positive p values the masses are start closer together at the bottom of the slinky and for negative p they start closer together towards the top.

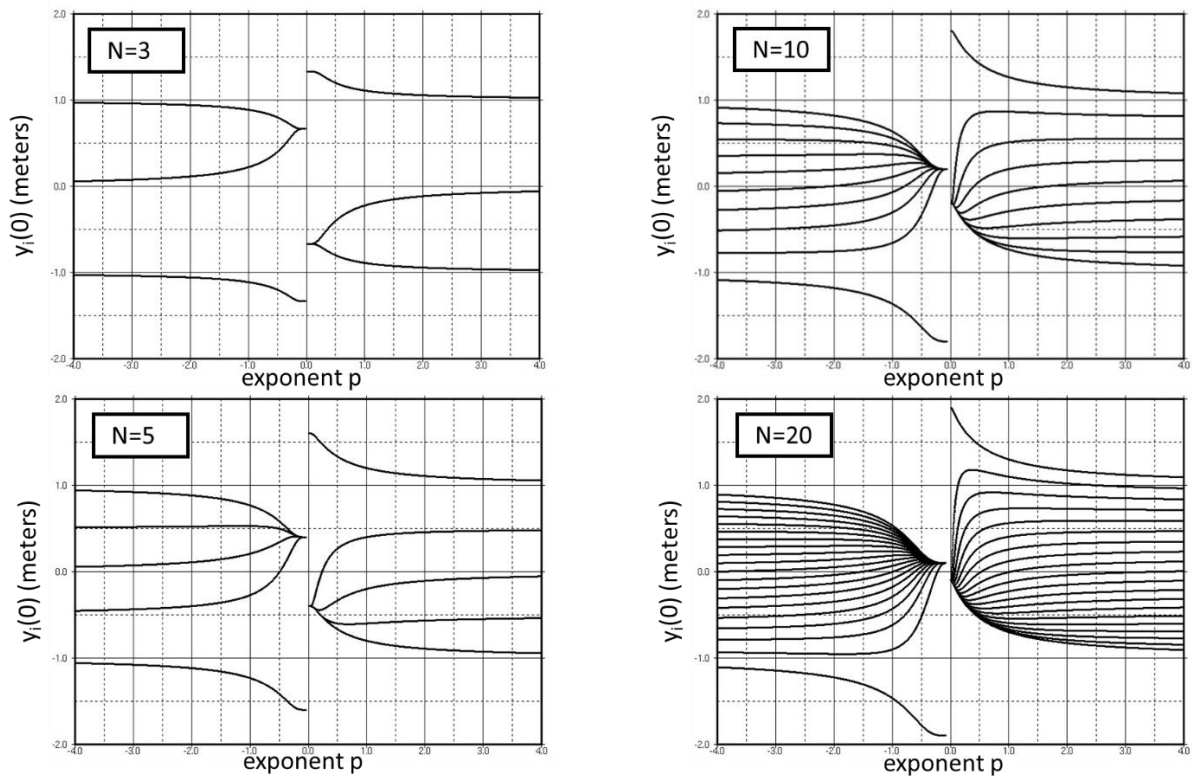

*Figure 2: Starting equilibrium positions of standard $L_S$ slinkys for several values of N. In this example, $L_S$=2.0 m and the masses are uniform.*

When p>0, the point masses are in a state of stable equilibrium at the starting time. Stable equilibrium means that the forces due to small fluctuations in the mass positions away from equilibrium tend to move the masses back towards the equilibrium position. For the p>0 slinky, equilibrium is achieved by simply holding the top mass stationary and allowing the slinky to hang freely. A very different situation occurs when p<0, because the masses are in a state of unstable equilibrium as the forces due to small fluctuations in positions away from equilibrium will move the masses further away from equilibrium. Achieving equilibrium with a p<0 slinky will require careful positioning of the masses and small errors will result in a breakdown of the slinky. For this reason, the p<0 slinky is not a fun toy.

There is a numerical instability near p=0 that we need to be aware of and avoid in our calculations. Suppose we consider the uniform mass problem and compute the ratio of the spacing between the upper two and lower two masses. Evaluating eq. (2) when the masses are uniform and N>2 we find

$$\text{Ratio} = \frac{y_N - y_{N-1}}{y_2 - y_1} = (N-1)^{1/p} \ . \tag{4}$$

If we have small absolute values of p and large values of N this ratio (or its reciprocal) will become larger than number of significant digits used for conventional numerical calculations (~16 decimal digits in a 64-bit IEEE floating point number). For example, when N=11 and p=0.1 this ratio is $10^{10}$! In practice, this instability is worse numerically when p<0 because the close clustering at the top of the slinky occurs near where the restoring forces become infinite and the lack of numerical precision is substantially amplified. When p>0, this close clustering occurs where the restoring force is zero. Programmer beware! Ways to avoid the instability are 1) put a minimum value on absolute value of p, or 2) carry out convergence tests and reject solutions not passing a desired threshold. The second method is used here and examples of this are shown in appendix B.

**Numerical Solutions**

With known starting positions, we are now ready to apply basic laws of mechanics to integrate for the time evolution of the slinky when the top is released. The force/mass of the acceleration $a_i = d^2 y_i/dt^2$ acting on the $i^{th}$ mass are:

$$a_i = -g - (y_i - y_{i-1})^p k/m_i + (y_{i+1} - y_i)^p k/m_i, \text{ when } 1 < i < N \tag{5a}$$

$$a_i = -g + (y_{i+1} - y_i)^p k/m_i \text{ when } i=1, \tag{5b}$$

$$a_i = -g - (y_i - y_{i-1})^p k/m_i, \text{ when } i=N. \tag{5c}$$

To solve this numerically we chose a sufficiently small step size $\Delta t$, and introduce the point mass velocity $v_i = dy_i/dt$ which starts out 0 in the equilibrium position at t=0. At each step in our numerical integration, the positions and velocities are updated using a simple first order integration method:

$$t \rightarrow t + \Delta t \tag{6a}$$

$$y_i \rightarrow y_i + v_i \Delta t \tag{6b}$$

$$v_i \rightarrow v_i + a_i \Delta t, \tag{6c}$$

where we explicitly keep track of the time variable t. Eqs (5) and (6) are then repeatedly iterated to a desired final time of interest. The numerical integration method shown in Eqs (6) is known as Euler's method and the accuracy of its solution depends on the step size, $\Delta t$. We also consider other integration methods such as 4$^{th}$-order Runge-Kutta[10] and in Appendix B discuss the pros and cons of these two approaches and selecting an appropriate step size.

For the p<0 case, we have to take care to properly handle the singularity that occurs at small displacements due to the divergent force term $k(y_i-y_{i-1})^p$. An effective way to handle this is to introduce a maximum allowed force, $f_{max}$, and replace these terms by $\max(f_{max}, k(y_i-y_{i-1})^p)$. For the simulations in this paper, a total slinky mass of 1kg is used and we set $f_{max}$=10000N which is more than 1000 times the combined forces due to gravity.

As the iterations progress, there will come a point where a pair of neighboring point masses comes into collision and care needs to be taken about how to detect and handle this circumstance. It straightforward to check whether a collision will occur in a given timestep. Prior to applying eq (6) we first test whether there is exists value of i such that $(y_{i+1}+ v_{i+1}\Delta t) \leq (y_i + v_i \Delta t)$. If such a value exists, there will be a collision between point masses i and i+1 during this time step.

There are three obvious choices for how to handle collisions when they occur:

1) Perfectly elastic collisions
2) Pass-through
3) Inelastic collisions (masses stick together).

Figure 3 illustrates these scenarios for a standard Hooke's law slinky having $L_S$=2.0 m, N=10, and p=1. In this example, the trajectories of 10 point masses vs time are computed and displayed with different colors together with the location of the center of mass of the system (black) which is the same in all three cases. Examining these three results and comparing with actual experiments with real slinkys (compare, for example, Fig. 3 in Cross and Wheatland[6]) the inelastic collision is the preferred solution and will be the approach taken in the remainder of this paper. Similar conclusions regarding the inelastic collisions were previously reported by Vanderbei[9] and Plaut et al[8].

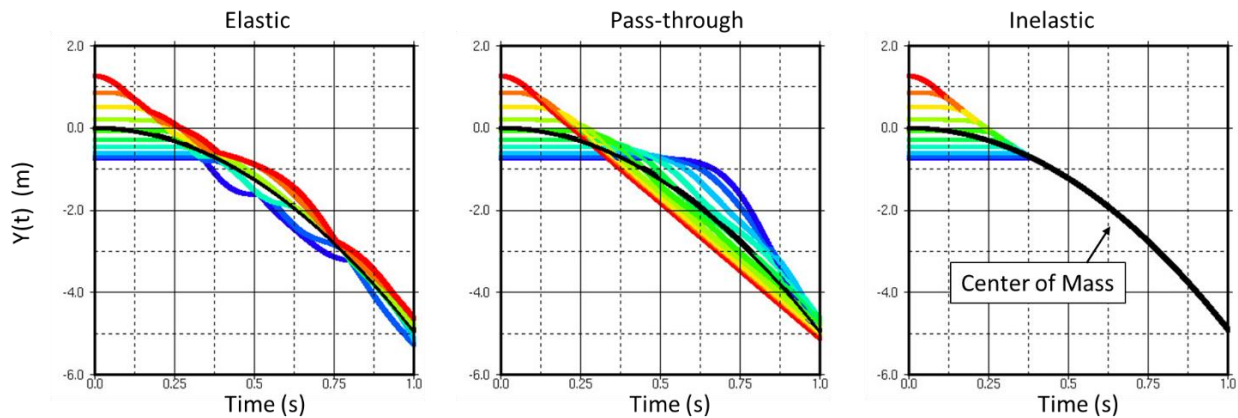

Figure 3: Comparison of Elastic, Pass-through, and inelastic collisions for standard slinkys with $L_S$=2.0m, N=10, p=1. Individual point mass trajectories are indicated by colored lines and the center of mass trajectory is in black.

Inelastic collisions are easily dealt with using numerical modeling. As before, we first detect whether a collision will occur between two masses i and (i+1) in a given timestep by checking whether $(y_{i+1}+ v_{i+1}\Delta t) \leq (y_i + v_i \Delta t)$. When such a pair is detected, it is a good practice to modify the step size, $\Delta t$, for this particular step so that the collision occurs exactly at the end of the step. For this step we define an updated step size,

$$\Delta t^* = (y_{i+1}-y_i)/(v_i-v_{i+1}), \qquad (7)$$

and iterate eqs. (6) using $\Delta t^*$. At the end of the step we apply the rules for an inelastic collision where particles i+1 and i merge into a single particle while conserving mass and momentum. We place the two merged particles at the $i^{th}$ position and as the $(i+1)^{th}$ mass has disappeared we need to make a few updates to the positions and velocities of the particles above the ith location. Capturing all these requirements we have:

$$v_i \rightarrow (mv_i + mv_{i+1})/(m_i + m_{i+1}) \qquad (8a)$$

$$m_i \rightarrow m_i + m_{i+1} \qquad (8b)$$

$$(y_{j+1}, m_{j+1}, v_{j+1}) \rightarrow (y_j, m_j, v_j) \text{ for } j=[i+1, N-1] \qquad (8c)$$

$$N \rightarrow N-1. \qquad (8d)$$

In the inelastic case, there is a unique collapse time and position where all the masses have merged into a single particle and we denote the time and position of this event by $t_C$ and $y_C$, respectively. After the collapse event occurs the equation of motion for the final point mass is $y_1(t)=½gt^2$ ($t \geq t_C$) because the center of mass of the system falls with an acceleration of g and after the collapse event the single remaining particle defines the center of mass. Note that if we change the starting length of the slinky, $L_S$, that $t_C$ and $y_C$ will scale proportionately by $(L_S)^{1/2}$ and $L_S$, respectively.

The movement of the bottom of the slinky at the time of collapse is $\Delta_{bottom}=y_1(0)-y_C$ and is a property we will examine in the next section to determine the conditions for whether the "bottom does not move" property holds.

**Modeling examples**

Figure 4 shows examples of applying these numerical solutions to Hooke's law slinkys (p=1) starting with values of N ranging from 2 through 9 and where the simulations stop at the collapse time. When there are just two point masses there is a significant downward motion (~ -0.2337 m) of the bottom mass at the time of collapse and as the number of number of starting point masses is increased the movement of the bottom mass becomes rapidly less. For example, when N=9, $\Delta_{bottom}$ ~ $-0.5004 \times 10^{-3}$ m which is quite negligible and impossible to observe given the resolution of the figure. This implies that the bottom-does-not-move property for Hooke's law slinkys results from the combined interaction of multiple particles.

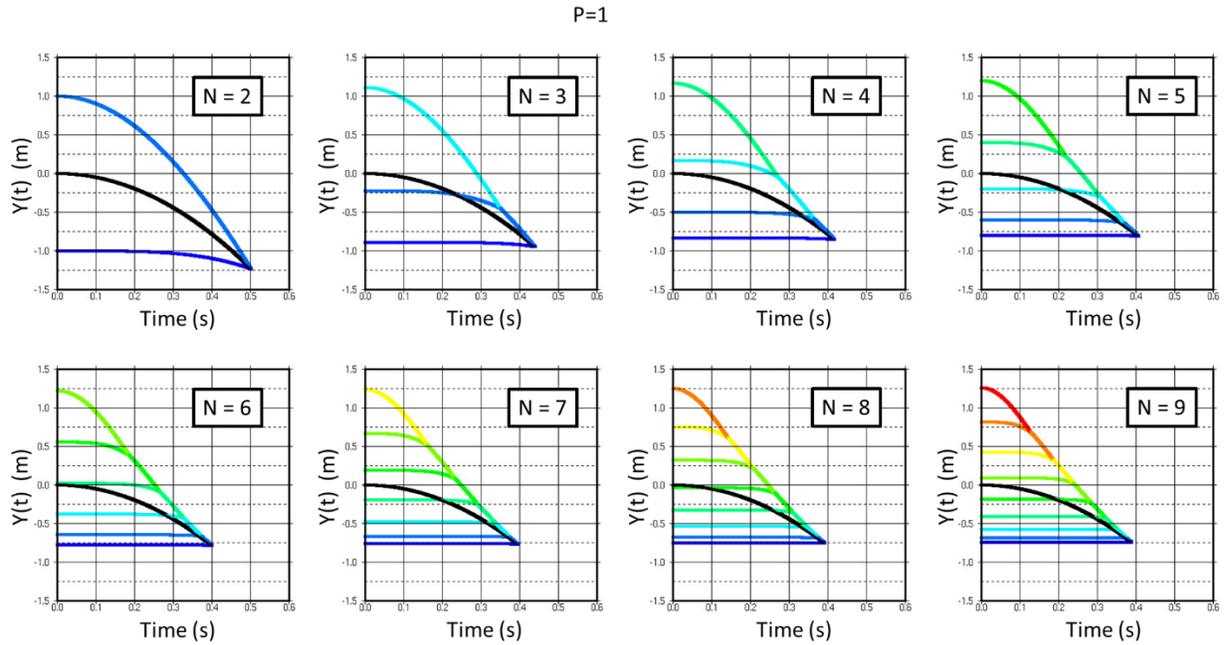

*Figure 4: Trajectories for standard LS Hooke's law (p=1) slinkys up until the collapse time where the number of starting masses, N, changes from 2 to 9. The center of mass trajectory is shown using a black curve and the colored curves are for the individual masses.*

To investigate what happens when we vary the slinky physics, lets first look at what happens when we change the starting mass distribution of the slinky. We consider 3 variations from the previous uniform mass distribution. The first is a top loaded slinky where $m_i = i*m$, the second is a bottom loaded slinky where $m_i = (N+1-i)*m$ and the third where the slinky masses are chosen such that $m_i = 0.001 + R_i$, where $R_i$ is numerically generated pseudorandom number chosen from a uniform distribution where $0 \leq R_i < 1$. The addition of the term .001 is to avoid particles of zero mass.

Figure 5 shows the trajectories of these different starting mass distributions of p=1 slinkys starting with N=9 particle masses. The examples from left to right correspond to the bottom loaded, random, uniform, and top loaded slinky mass distributions. Beneath the display of the particle trajectories are the starting mass distributions. We observe negligible movement at the bottom of the slinky for these very different starting mass distributions. The bottom-does-not-move property holds for p=1 slinkys with varying mass distributions!

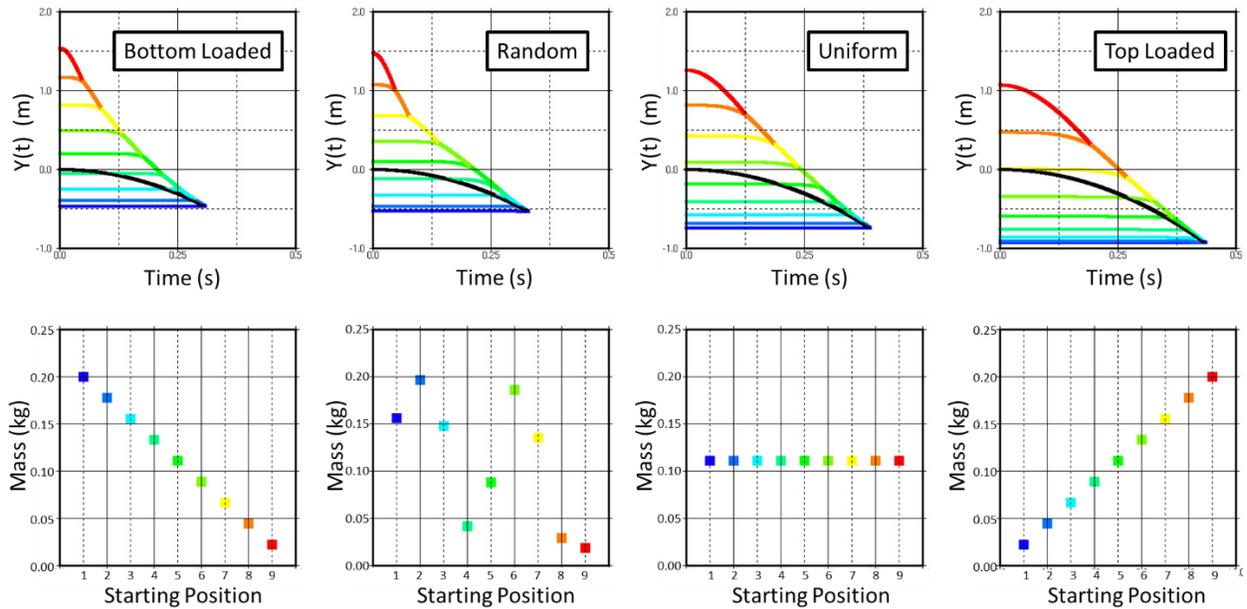

*Figure 5: Top - Comparison of particle trajectories up through collapse time for standard $L_S$=2.0 m Hooke's law (p=1) slinkys starting with 9 particles. Four different starting mass distributions are illustrated. Bottom of the figure shows the starting particle masses for each distribution.*

We now look at what happens when we vary the exponent p. Particle trajectories are computed for a range of cases where -4 ≤ p ≤ 4, and N ranges from 2 to 50. In Appendix (B) a detailed examination of the convergence of these solutions are presented that show accuracy at the $10^{-3}$ m level with some (p,N) values needing removal due to numerical instabilities as described above concerning eq. (4). The movement of the slinky bottom, $\Delta_{bottom}$, was computed at the time of final collapse for each value of p and N that was simulated and the results are displayed in figure 6.

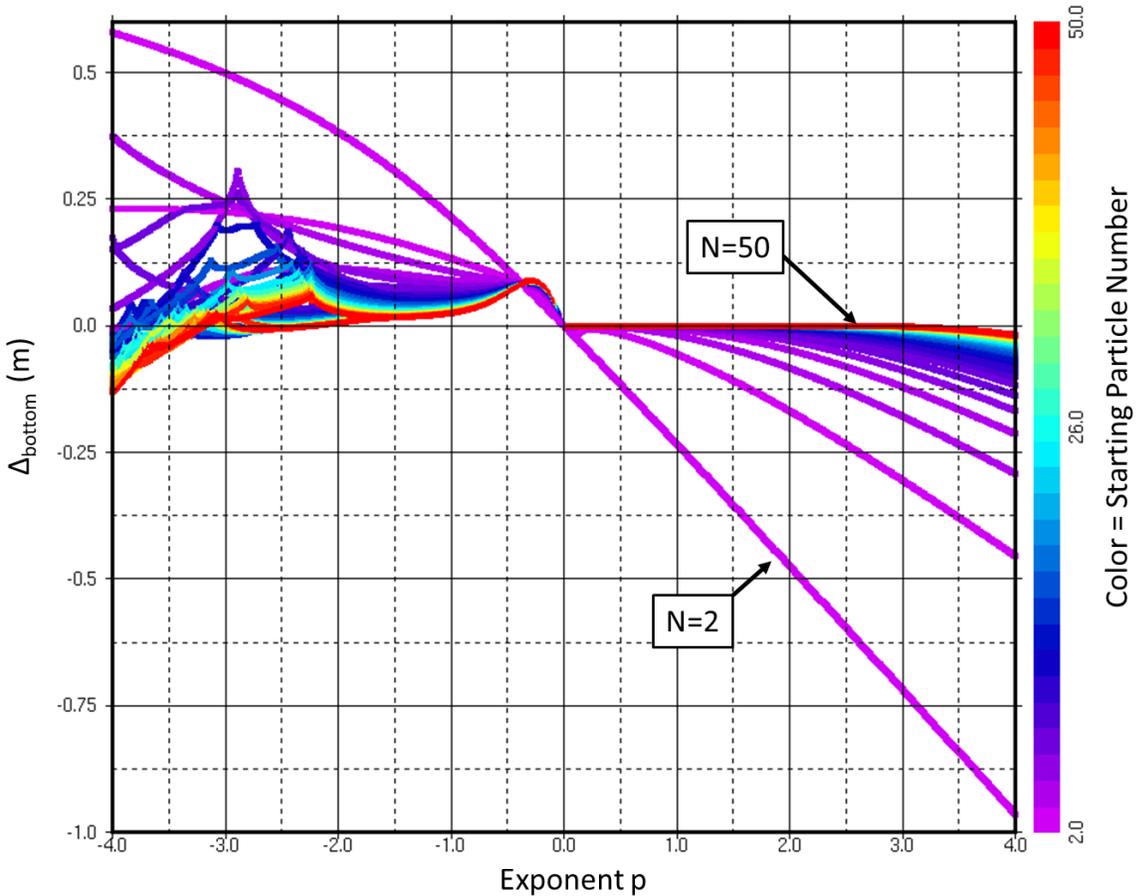

*Figure 6: Movement of the slinky bottom at the time of collapse (vertical axis) versus the p-exponent value (horizontal axis). Color indicates the starting number of point masses and ranges from 2 (purple) to 50 (red).*

When N=2, there is significant downward motion of the bottom mass for positive p and upward motion for negative p. In Appendix A, the N=2 case is examined in more detail and an analytic solution for the collapse time and position is derived for all p values.

For positive p values it is observed that the bottom-does-not-move property always holds provided that the starting number of particles is sufficiently large. We can see that the movement of the bottom mass quickly tends towards zero from below as the number N is increased. As p increases, a larger number of particles are required to keep the movement of the slinky bottom near zero.

For the negative p values, there are many new phenomena occurring that are not observed when p is positive. There are a number of cusps developing in the solutions with the most notable occurring when N=6, p ~ -2.89. There is also a region in the range -3 < p < -2 where as N gets large the solutions bunch up into two different sets. The upper group correspond to values of N that are even and the lower group corresponds to values of N that are odd. The reason for this even/odd selection has to do with the fact the at these p values the slinky starts collapsing across its whole length with neighbors pairing and inelastically colliding with one another. A slinky starting with an even N can do this more efficiently than that starting with an odd N. Some example trajectories comparing odd versus even starting N values are shown in Figure 7 for a p=-2.5 slinky. Note how when N is even the particles can pair exactly but that when N is odd there is always an unpaired mass that delays the point of final collapse.

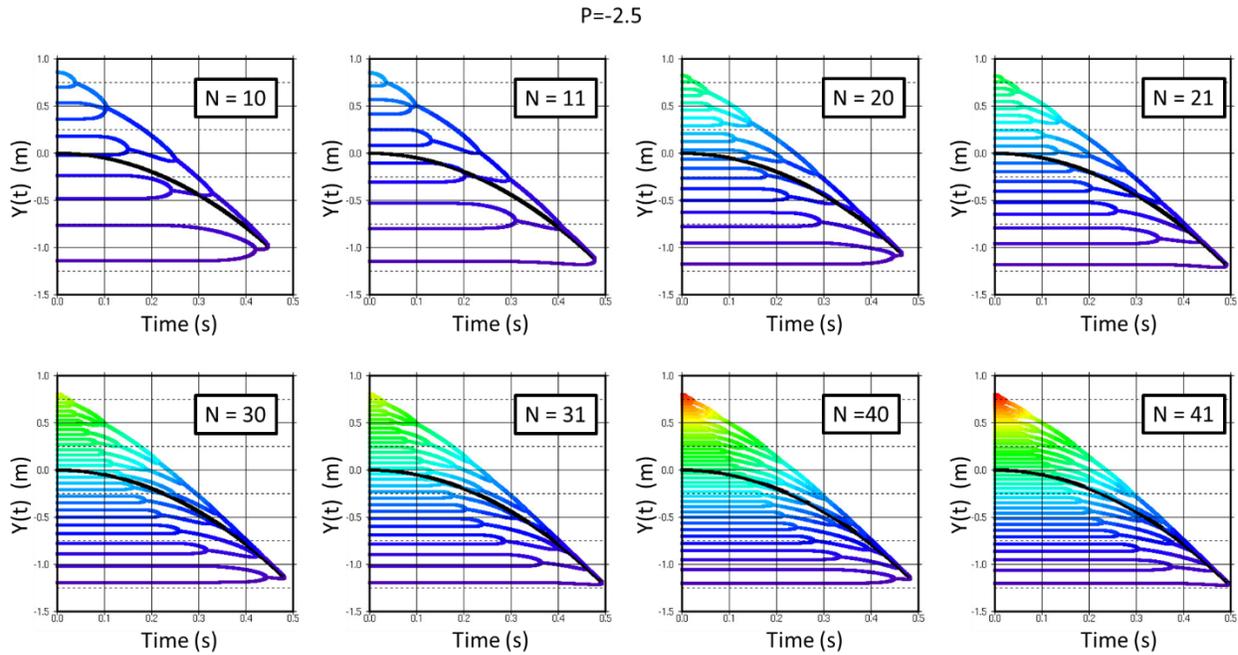

*Figure 7: Trajectories for standard $L_S$=2.0 m, p=-2.5 slinkys comparing even vs odd number of starting masses. The center of mass trajectory is shown in black and the colored trajectories represent the individual particles.*

Figure 8 shows the calculation of $\Delta_{bottom}$ for a p=-2.5 slinky where the number of point masses ranges from 2 to 100. We can clearly see the separation between even and odd numbers of point masses. There is also a subtle pattern in these results that also depends on N modulo 4.

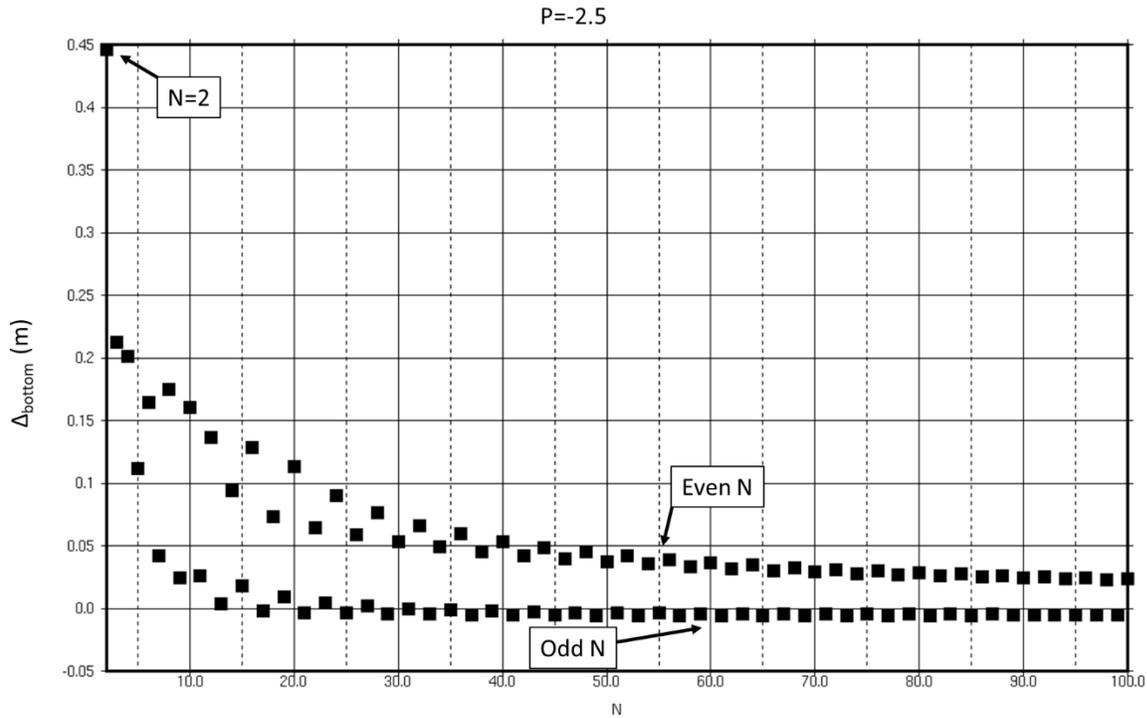

*Figure 8: Movement of the slinky bottom at the time of collapse (vertical axis) vs the number of point masses in the model for a standard slinky with $L_S=2.0m$ and p=-2.5.*

For a finite number of particles such as those modeled here (N≤50), the bottom-does-not-move property is not valid when p<0. Can we answer the question about what happens as the number of particles becomes infinite? This is a difficult numerical question to answer for several reasons. The main one has to do with complications that arise in many-particle systems with force laws such as gravity (p=-2) where it is well known that high sensitivity to initial positions leads to chaos and unpredictability of trajectories. For example, Mikkola and Hietarinta [11](1989) describe 1-dimensional solutions to the 3-body gravitational force problem (with pass-through interactions) that illustrate the rich complexity that such problems entail. Although the initial conditions in the falling slinky problem are known precisely, numerical computations involve finite numbers of digits, numerically rounding, and will not succeed in answering this question for arbitrarily large N. In appendix B, it is shown that we are already starting to encounter into these numerical issues for the case p=-4, N-50.

**Conclusions**

Real slinkys released to fall from a stretched equilibrium position have the interesting property that the bottom does not move until the collapsing upper layers reach the bottom. In this paper, we examined numerical solutions of the generalized falling slinky problem to investigate whether the bottom-does-not-move property also holds for non-harmonic slinkys such as deviating the restoring force equations from Hooke's law or using non uniform mass distributions.

The numerical models presented here consisted of slinkys made up of N point masses interacting with restoring forces, F, of the type $F = kx^p$, where x represents the spring displacement. The most important property of the slinky that controls its dynamic behavior is its equilibrium length, $L_S$, and so we defined standard length=$L_S$ slinkys with starting center of mass located at 0. Having standard $L_S$ models makes it easier to compare different cases of mass, p, and particle number, N.

After the slinky is released from its equilibrium position, the particles collide with one another and inelastic collisions between the point masses are used for these collisions. With this model, there is a unique collapse time, $t_c$, and position, $y_c$, defined when all the masses have merged into a single particle. For the standard slinky, $y_c=½gt_c^2$, and beyond the collapse time the equation of motion for the final point mass with time follows $½gt^2$ ($t≥t_C$). If we change the starting length of the slinky, $L_S$, the values of $t_C$ and $y_C$ will scale proportionately by $(L_S)^{1/2}$ and $L_S$, respectively.

Our investigation showed that for small values of N, there is substantial motion of the slinky bottom for absolute values of p that are non-zero. The method applied to the Hooke's law slinky (p=1) with uniform mass showed that as the number of particles N is increased beyond 2, that the bottom-does-not-move property soon takes hold and by the time N>9, it is difficult to observe any movement of the slinky bottom until the collapse time.

Changing the mass distribution of the Hooke's law slinky did not change the bottom-does-not-move property. We looked at top-loaded, bottom-loaded, and random mass distributions and saw that for large N that the bottom-does-not-move property still held.

For non-Hooke's law forces it is found the when $p > 0$, the bottom-doesn't-move property holds for large N, but when $p < 0$, the model shows complicated collapse patterns that in some cases depend on whether the number of modeled masses is even or odd and this was attributed to the observation that for a range of negative p values that the particles collapsed in neighboring pairs; a processes that is efficiently completed with an even number of particles but not with an odd number.

One difference between the generalized positive and negative p slinkys is that the former start from a state of stable equilibrium and the latter from a state of unstable equilibrium. Although there are many negative-p force laws of great interest to physicists such as gravity and electrostatic (p=-2) and magnetic dipole (p=-3) assembling these into a linear chain as modelled here would be difficult in practice. It is a well-known problem in physics that the solution to the many-body problem with negative p forces are inherently chaotic and extremely sensitive to initial conditions. This leads to a limitation of the numerical methods that cannot accurately solve the dynamical system for large N to investigate what happens at N=∞.

**Appendix A: Solution for N=2 with uniform masses.**

In this appendix we look at the solution for the movement of the bottom mass at the moment of collapse when N=2 and setting the masses identical so that $m_i=m$. In this case, eqs. (5) reduces to two equations:

$$a_1 = -g + (y_2-y_1)^p k/m \qquad (A1)$$

$$a_2 = -g - (y_2-y_1)^p k/m \qquad (A2)$$

Adding A1 and A2 yields an equation for the center of mass $y_{cm}=(y_1+y_2)/2$, and its acceleration $a_{cm}$,

$$a_{cm} = -g,$$

with solution $y_{cm} = \frac{1}{2}gt^2$.

Subtracting eq. (A1) from eq. (A2) yields an equation for the (positive) difference $y=y_2-y_1$, and its acceleration $a=d^2y/dt^2$

$$d^2y/dt^2 = -2y^p k/m. \qquad (A3)$$

The initial condition that the starting length=$L_s$ at time t=0 requires that $k=mg/L_s^p$ so that eq. (A3) can be written as:

$$d^2y/dt^2 + \alpha y^p = 0, \qquad (A4)$$

where $\alpha = 2g/L_s^p$.

After multiplying A4 by $\frac{dy}{dt}$ and introducing the potential $V(y) = \int \alpha y^p dy$, this becomes:

$$\frac{dy}{dt}\frac{d^2y}{dt^2} + \frac{dy}{dt}\frac{dV}{dy} = \frac{d}{dt}\left[\frac{1}{2}\left(\frac{dy}{dt}\right)^2 + V\right] = 0. \qquad (A5)$$

The term inside the brackets [] (corresponding to energy) has a zero time-derivative and is therefore a constant of the system. At time $t=0$, the term $\frac{dy}{dt}=0$, and the total energy has the value $V(L_s)$. After some algebraic manipulations we can write this as:

$$dt = -\frac{dy}{\sqrt{2(V(L_s)-V(y))}}, \tag{A6}$$

where it should be observed that the negative square root of $\left(\frac{dy}{dt}\right)^2$ is chosen because y is a decreasing function of t and at $t = 0$, $y = L_s$. The collision time between the two masses occurs when $y = 0$, and we can find this by integration both sides of Eq.(A6):

$$t_c = \int_0^{L_s} \frac{dy}{\sqrt{2(V(L_s)-V(y))}} \tag{A7}$$

For $p \neq -1, V(y) = \frac{\alpha y^{p+1}}{p+1}$ and when $p = -1, V(y) = \alpha \ln(y)$ and so we need to examine these two cases separately. Starting with the first case, Eq.(A7) becomes:

$$t_c = \sqrt{\frac{L_s}{g}} f(p), \tag{A8}$$

where

$$f(p) = \frac{\sqrt{p+1}}{2} \int_0^1 \frac{dz}{\sqrt{1-z^{p+1}}}. \qquad (p>-1) \tag{A9a}$$

$$f(p) = \frac{\sqrt{|p+1|}}{2} \int_0^1 \frac{dz}{\sqrt{z^{p+1}-1}}. \qquad (p<-1) \tag{A9b}$$

As a side note eq. (A9a) can be evaluated[12] in terms of the gamma function:

$$f(p) = \frac{\sqrt{\pi}}{2\sqrt{p+1}} \frac{\Gamma(\frac{1}{p+1})}{\Gamma(\frac{1}{p+1}+\frac{1}{2})}, \qquad (p>-1) \tag{A10}$$

The difference between the location of the bottom mass at the time of collision ($-\frac{1}{2}gt_{collision}^2$) and its starting position ($-L_s/2$) is:

$$\Delta_{Bottom} = -\frac{1}{2}L_s(f^2(p) - 1). \tag{A11}$$

The function f(p) does not have simple analytical solutions for arbitrary values of p. One does exist for the harmonic value of $p = 1$, in which case $t_{collision} = \frac{\sqrt{2}\pi}{4}\sqrt{\frac{L_s}{g}}$. The difference between the final location of the bottom mass ($-\frac{1}{2}gt_{collision}^2$) and its starting position ($-L_s/2$) is:

$$\Delta_{Bottom} = -L_s\left(\frac{\pi^2}{16} - \frac{1}{2}\right) \cong -0.1168 L_s, \tag{A12}$$

Which equals ~-0.2337 m when $L_s = 2.0$ m.

For the second case, where $p = -1$, eq.(A7) becomes:

$$t_{collision} = \frac{1}{\sqrt{2\alpha}} \int_0^{L_s} \frac{dy}{\sqrt{-\ln\left(\frac{y}{L_s}\right)}}. \tag{A13}$$

This is solvable by setting $y = L_s e^{-z^2}, dy = -2zL_s e^{-z^2}$ to obtain

$$t_{collision} = \frac{2L_s}{\sqrt{2\alpha}} \int_0^\infty e^{-z^2} dz = \frac{\sqrt{\pi}}{2}\sqrt{\frac{L_s}{g}}. \tag{A14}$$

For the case $p = -1$, we find

$$\Delta_{Bottom} = -L_s\left(\frac{\pi}{8} - \frac{1}{2}\right) \cong +0.1073 L_s, \tag{A12}$$

which equals 0.2146 m when $L_s = 2.0$ m.

Note that when p=0, there is a simple solution for the case when N=2. The upward spring force on the bottom mass is mg (exactly countering gravity) and the downward spring force on the upper mass is also mg which combines with the gravitation force to produce a downwards acceleration of 2g once the top mass is released. So in this case, the bottom mass will not move at all until the collapse time so that:

$$t_{collision} = \sqrt{\frac{L_s}{g}}, \tag{A13}$$

$$\Delta_{Bottom} = 0. \tag{A14}$$

**Appendix B: Analysis of the convergence of the numerical solution**

In this appendix we look at the accuracy of the numerical solutions and carry out tests to ensure the step size, Δt, used in the iteration of eqs. (6) is sufficiently small enough to reach a solution of desired accuracy (stated as <1.0 mm for final collapse position, $y_C$).

The best confidence test is to compare numerical solutions with known exact results. Although there are no general analytical solutions that we are aware of when N>2, we do know from the preceding analysis that at the time of collapse, $t_C$, that the position of the center of mass should occur exactly at ½$gt_C^2$. A good first test is to compare our calculated center of mass with the known center of mass to check if there are any systematic errors and how these errors depend on the step size. We should also look at final results such as $\Delta_{Bottom}$ and see how depends on the step size.

As mentioned previously, the integration method of eqs. (6) is called the Euler method and we will compare this integration method to one using a 4th order Runge-Kutta[10] integration (RK4). The RK4 method involves calculating the RHS of eqs. (5) at four different locations and is computationally is 4 times as expensive as an Euler step using the same overall step size.

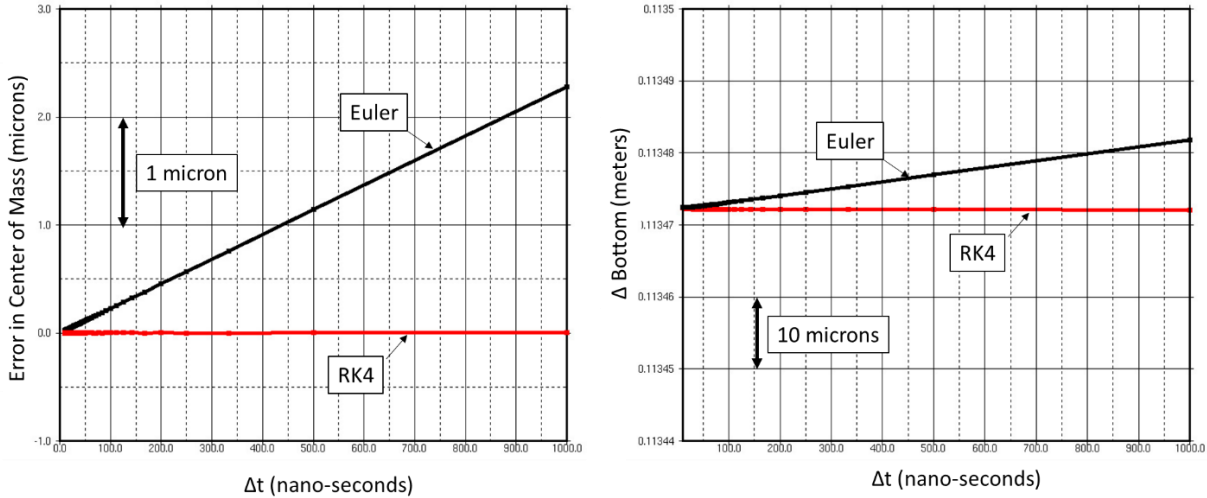

*Figure B1: Comparison between errors in final center of mass locations for the Euler and RK4 method vs step size (left). Comparison between the movement of the slinky bottom, $\Delta_{Bottom}$, vs step size for the Euler and RK4 method (right). These results are for a standard slinky with p=-2.5, N=20, and $L_S$=2.0 m.*

On the left side of figure (B1) is a plot of the step size vs error in the computed center of mass location in microns ($10^{-6}$ m) defined by $Error_{CM}=10^6(y_c-\tfrac{1}{2}gt_c^2)$ for the Euler integration method (black) and the RK4 method (red) for a slinky with $L_S$=2, p=-2.5, N=20. On the right side is the computed final $\Delta_{Bottom}$ position vs stepsize. There are a many things worth pointing out in these displays. In terms of errors for this particular value of p and N, both the Euler and RK4 method have low errors in the center of mass (at the level of a few microns). As a function of step size this error changes linearly and the Euler and RK4 results converge at nearly the same hypothetical limit where Δt=0. The same overall details are observed on the right hand side plot that looks at the results for $\Delta_{Bottom}$ vs step size, although in this case the errors are an order of magnitude larger but still in the acceptable range.

The errors in the center of mass at the collapse point are used to detect the numerical instabilities that arise when the p values are near-zero as detailed in eq. (4). For example, figure B2 shows a map in the (p,N) space of the $Error_{CM}$ using the Euler integration with a step size of 125 nanoseconds. The color bar indicates the value of the error and is red when the error ≥ 100 microns (note that all the "purple" errors are less than 0.5 microns and show up as a single color). These errors are easily identified with this QC

and are removed in the final results such as those shown in Fig. (6)

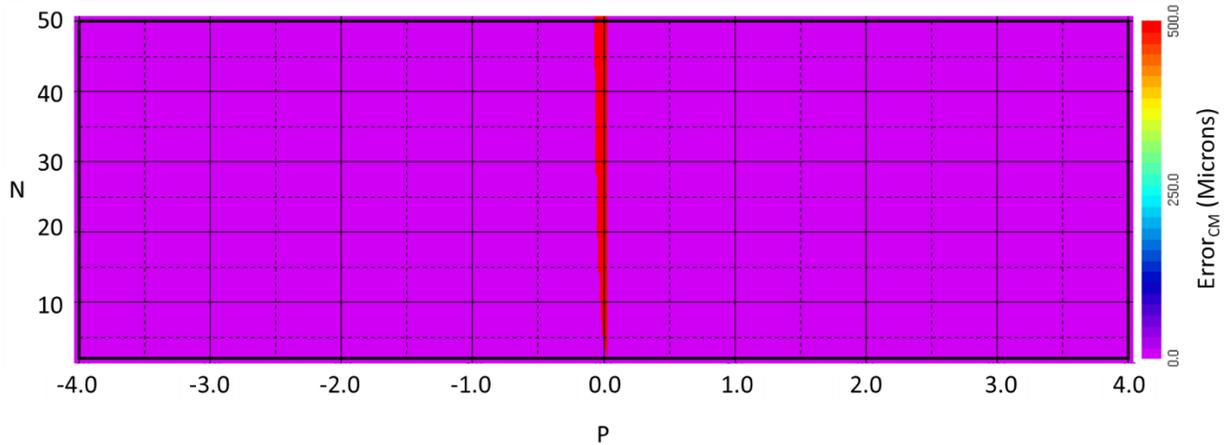

*Figure B2: Error in the center of mass location at the collapse time for all values of p and N. Red points correspond to large errors due to the numerical instability pointed out in eq. (4).*

The RK4 method has less error and variability than the Euler method but it comes at the price of costing 4x more computationally. Some advantages to the Euler method are 1) it is easier for students to code and incorporate into their own programs, and 2) during steps with a collision an exact step size can be chosen as outlined in eq. (7). Nevertheless, motivated students should try out the RK4 method and other integration schemes to learn the pros and cons of these methods.

As final error test, we calculate $\Delta_{Bottom}$ and look at how this quantity depends on the stepsize, $\Delta t$, used to generate it. One estimate of the error is to calculate the absolute difference between $\Delta_{Bottom}$ calculated at different step sizes. For our purposes we will compare step size $\Delta t$ with step size $\Delta t/2$ and define

$$\text{Error}_1(\Delta t) = | \Delta_{Bottom}(\Delta t) - \Delta_{Bottom}(\Delta t/2) |, \tag{B1}$$

where we introduced the notation $\Delta_{Bottom}(\Delta t)$ that corresponds to the functional dependence of $\Delta_{Bottom}$ and $\Delta t$.

Figure B3 shows values of $\text{Error}_1$ in eq. (B1) calculated at several step sizes for both the Euler (Top) and RK4 (bottom) methods. The color bar in this figure has a maximum error of 100 microns. For the Euler method, we can see that the values of $\text{Error}_1$ decrease as the step size is made smaller. By the time we have reached a step size of 125 nanoseconds this error falls below ~ 100 microns. There is a persistent trouble spot near the upper left corner of these (p=-4, N=50) that keeps large errors even at the smallest step size shown. For the RK4 method, the $\text{Error}_1$ quantity is much smaller than it is for the Euler method – supporting that this is a worthwhile integration method. Nevertheless, the upper left corner is also a problem spot for the RK4 method.

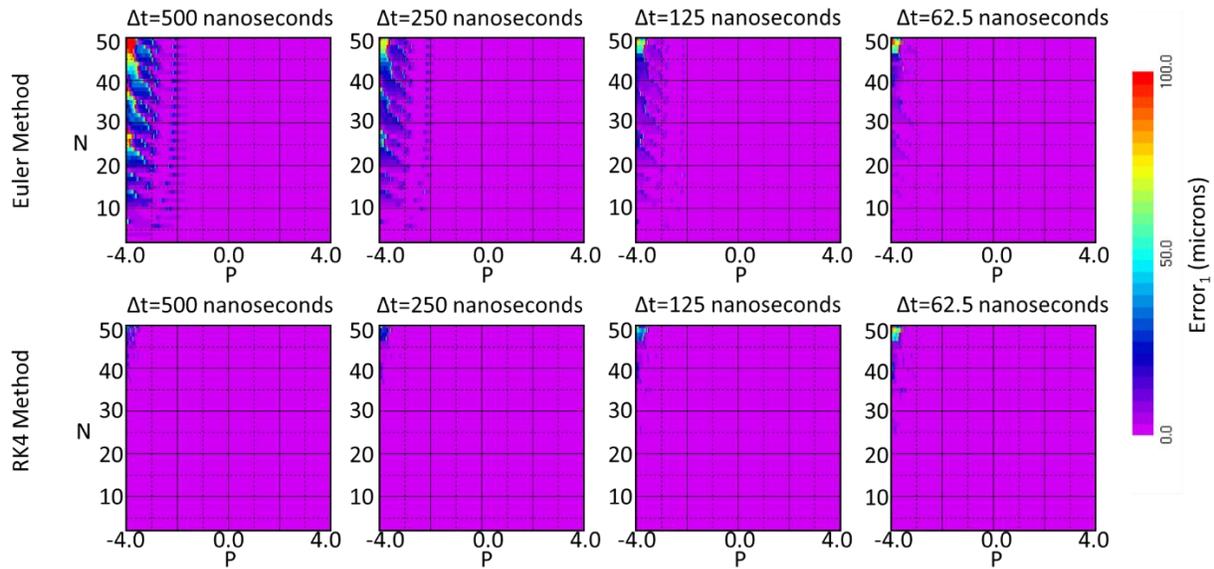

*Figure B3: Computed values of the term Error1, eq. (B1), calculated for several different step sizes using the Euler method (Top) and RK4 method (Bottom).*

To illustrate the problem occurring in the upper left corner near p=-4, N=50, figure B4 shows that values of $\Delta_{Bottom}$ calculated versus step size for the Euler and RK4 methods. The left side of the figure is generated using conventional "double precision" floating point numbers (64-bit) and the exercise is repeated on the right using higher precision floating point numbers "quad precision" (128-bit). Note that for the 64-bit floating point numbers that as step size moves towards the hypothetical limit of Δt=0 that both the Euler and RK4 methods do not converge linearly to a result. For the 128-bit calculation, these results are linear. Why is this? The answer is that we are dealing with a many-particle dynamical system with high sensitivity to starting initial positions and truncation errors that take place in numerical solutions are now leading to problems like those on the left side of fig. (B4).

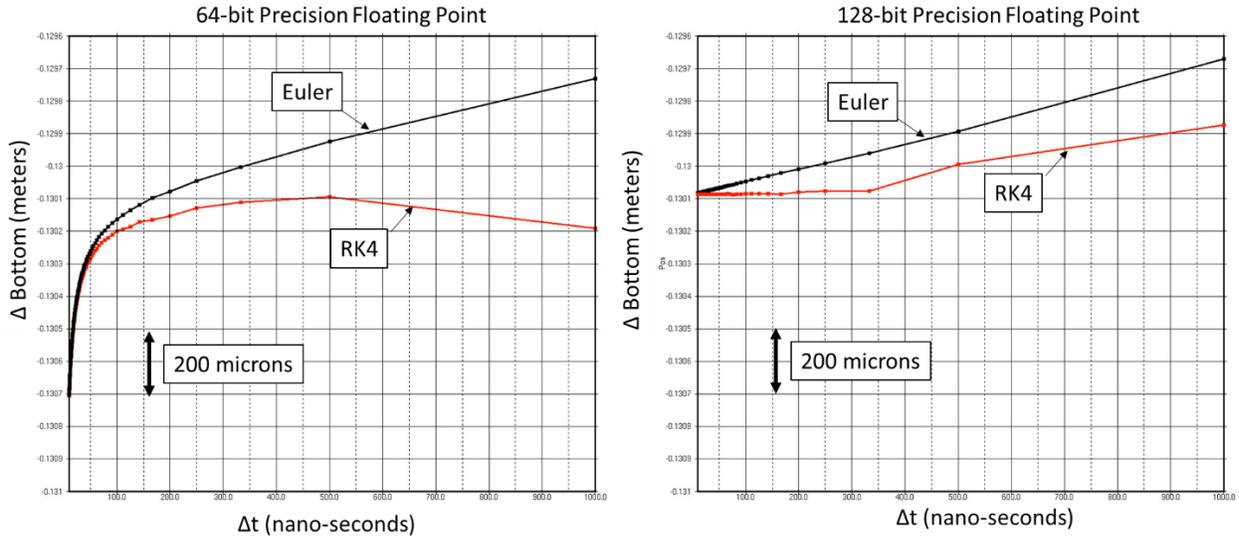

*Figure B4: Comparison between the movement of the slinky bottom, $\Delta_{Bottom}$, vs step size for the Euler and RK4 method. The left-hand plot is calculated using 64-bit floating point numbers and the right-hand plot is calculated using 128-bit floating point numbers. These results are for a standard slinky with p=-4.0, N=50, and $L_S$=2.0 m.*